\documentclass[aps]{revtex4-2} 
\usepackage{graphicx,grffile,wrapfig,overpic}
\usepackage{amsmath,amssymb,mathrsfs,eufrak,eucal}
\usepackage{braket,mathtools,cancel,bm,siunitx}
\usepackage[colorlinks,linkcolor=mediumtealblue,citecolor=Ao(English),bookmarksopenlevel=4]{hyperref}
\usepackage[usenames]{color}
\usepackage{ulem}
\usepackage{tikz,tikz-cd}
\usepackage[small,bf]{caption}
\usepackage{subcaption}
\usepackage{dcolumn}
\newcommand{\dmeasure}[2]{\mathrm{d}^{#1} #2 \,}
\newcommand{\tr}{\mathrm{tr}\,}
\newcommand{\psibar}{\bar{\psi}}
\newcommand{\Gammaf}[1]{\Gamma\left(#1\right)}
\newcommand{\PolyLog}[2]{\mathrm{Li}_{#1}\left(#2\right)}
\newcommand{\zetaf}[1]{ \zeta \left( #1 \right)}

\newcommand{\ie}{\textit{i}.\textit{e}., }
\newcommand{\eg}{\textit{e}.\textit{g}.\ }

\definecolor{blue}{rgb}{0.0, 0.0, 1.0}
\definecolor{Ao(English)}{rgb}{0.0, 0.5, 0.0}
\definecolor{mediumtealblue}{rgb}{0.0, 0.33, 0.71}
\graphicspath{{graph/}}

\begin{document}

\title{Super Restoration of Chiral Symmetry in Massive Four-Fermion Interaction Models}

\author{Tomohiro Inagaki}
 \email{inagaki@hiroshima-u.ac.jp}
 \affiliation{Core of Research for the Energetic Universe, Hiroshima University, Higashi-Hiroshima, 739-8526, Japan}
 \affiliation{Information Media Center, Hiroshima University, Higashi-Hiroshima, 739-8521, Japan}

 \author{Daiji Kimura}
 \email{kimurad@ube-k.ac.jp}
 \affiliation{National Institute of Technology, Ube College, Ube, 755-8555, Japan}

 \author{Hiromu Shimoji}
 \email{h-shimoji@hiroshima-u.ac.jp}
 \affiliation{Information Media Center, Hiroshima University, Higashi-Hiroshima, 739-8521, Japan}

\date{\today}

\begin{abstract}
 The chiral symmetry is explicitly and spontaneously broken in a strongly interacting massive fermionic system.
 We study the chiral symmetry restoration in massive four-fermion interaction models with increasing temperature and chemical potential.
 At high temperature and large chemical potential, we find the boundaries where the spontaneously broken chiral symmetry can be fully restored in the massive Gross--Neveu model.
 We call the phenomenon super restoration.
 The phase boundary is obtained analytically and numerically.
 In the massive Nambu--Jona-Lasinio model, it was found that whether super restoration occurs depends on regularizations.
 We also evaluate the behavior of the dynamical mass and show the super restoration boundaries on the ordinary phase diagrams.
\end{abstract}

\maketitle

\section{Introduction}\label{sec:introduction}
Chiral symmetry is a fundamental property of elementary particles.
In the Quantum chromodynamics (QCD), quarks and gluons are confined into hadrons, which is related to the chiral symmetry breaking.
The QCD Lagrangian for the light quarks possesses an approximate chiral symmetry due to non-vanishing current quark masses.
At low energy scale, the constituent quark masses are dynamically generated by the spontaneous chiral symmetry breaking.
It is expected that the thermal effect restores the broken chiral symmetry at high temperature and/or large chemical potential.
One of interesting topics in QCD is to investigate chiral symmetry breaking and restoration.

We can not avoid the non-perturbative effect of QCD in order to study the chiral symmetry breaking.
One of possible procedures is to consider the phenomena in a low energy effective model of QCD.
The Gross--Neveu (GN) model \cite{Gross:1974jv} is often used to investigate the phase structure of the chiral symmetry in extreme conditions.
It is a renormalizable model with a four-fermion interaction since the model is defined in a two-dimensional spacetime.
In the original GN model, the discrete $\mathbb{Z}_2$ chiral symmetry prohibits mass terms.
The four-fermion interaction induces non-vanishing expectation value for the composite operator constructed by the fermion and the anti-fermion, and the chiral symmetry is broken spontaneously.
For $2\leq D <4$, four-fermion interaction models are renormalizable in a sense of the $1/N_c$ expansion and possess an ultraviolet-stable fixed point \cite{Eguchi:1977kh, Shizuya:1979bv, Rosenstein:1988pt}.
The broken chiral symmetry is restored at high temperature and/or large chemical potential and the phase boundary for the massless model has been analytically and numerically shown in Ref.~\cite{Inagaki:1994ec}.

In the massive GN model, the mass term breaks the chiral symmetry explicitly, and it is expected to avoid the no-go theorem \cite{Mermin:1966fe,Hohenberg:1967zz,Coleman:1973ci} with the mass term even with the axial interaction term at zero temperature.
The Nambu--Jona-Lasinio (NJL) model \cite{Nambu:1961tp, Nambu:1961fr} is a four-fermion interaction model with the current quark mass terms and considered as one of the low energy effective models of QCD.
The NJL type models often used in the analysis of phase diagrams at a finite temperature, $T$, and chemical potential, $\mu$.
Since the model contains the four-fermion interaction and is not renormalizable in the four-dimensional space-time, the model prediction may depend on the regularization method.
One introduces a cutoff scale to remove the UV divergence in the fermion loops.
The three-dimensional momentum cutoff scheme is often adopted (for reviews, \cite{Vogl:1991qt, Klevansky:1992qe, Hatsuda:1994pi, Buballa:2003qv}) to investigate the phase diagrams.
Other methods are also considered in the model, the Pauli--Villars \cite{Kahana:1992jm, Osipov:2004bj, Moreira:2010bx, Kraatz:2022xya}, the Fock--Schwinger proper-time \cite{Suganuma:1990nn, Klimenko:1990rh, Klimenko:1991he, Gusynin:1994re, Inagaki:2003yi, Cui:2014hya, Zhang:2016zto, Cui:2017ilj, Li:2017xlb, Wang:2018qyq, Cui:2018gkg}, and the dimensional regularization \cite{Krewald:1991tz, Jafarov:2003pe, Jafarov:2004jw, Inagaki:2007dq, Inagaki:2011uj, Inagaki:2012re, Liu:2014vha}.
Some papers comprehensively study these differences in the regularizations \cite{Kohyama:2015hix, Kohyama:2016fif}.

The paper is organized as follows: in Sec.~\ref{sec:model}, the massive GN model is introduced.
We renormalize the four-fermion coupling and mass parameter to obtain a well-defined effective potential at a finite temperature and chemical potential.
In Sec.~\ref{sec:super-restoration}, we derive the equation for the condition of super restoration from the gap equation.
We evaluate the behavior of the dynamical mass, and draw the phase boundary of the super restoration and the boundary of the chiral condensate on a $\mu$-$T$ plane.
In Sec~\ref{sec:two-flavor-njl}, we also compute the same physical quantities in the two-flavor NJL model with two different regularizations.
Finally, summary and discussions are given in Sec.~\ref{sec:summary}.

\section{Gross--Neveu Model}\label{sec:model}
We briefly introduce the action of the massive GN model on the $D$-dimensional spacetime ($2 \leq D < 4$),
\begin{equation}
 S = \int\dmeasure{D}{x} \left[ \psibar(x) \left( i\gamma^\mu\partial_\mu - m_0\right) \psi(x) + \frac{\lambda_0}{2N} \left(\psibar(x) \psi(x)\right)^2\right],
  \label{eq:GN:action1}
\end{equation}
where $m_0$ and $\lambda_0$ are a bare mass parameter and a bare coupling of a four-fermion interaction, respectively.
$N$ denotes the number of copies of fermions $\psi(x)$.
In the massless case, $m_0 = 0$, this model enjoys the discrete chiral $\mathbb{Z}_2$ symmetry.

Introducing an auxiliary field, $\tilde{\sigma}(x)$, to this action, we obtain a redefined action,
\begin{equation}
 S_a = \int\dmeasure{D}{x} \left[ \psibar(x) \left( i\gamma^\mu\partial_\mu - \tilde{\sigma}(x) \right) \psi(x) - \frac{N}{2\lambda_0} \left(\tilde{\sigma}(x)^2 - 2 m_0 \tilde{\sigma}(x) \right)\right].
  \label{eq:GN:action2}
\end{equation}
In this expression \eqref{eq:GN:action2}, the constant term, $N m_0^2/(2 \lambda_0)$, is dropped  because the term does not affect physical quantities.
The original action \eqref{eq:GN:action1} is reproduced by substituting the solution of the equation of motion, $\tilde{\sigma}(x) = - \lambda_0 \psibar(x)\psi(x)/N +m_0$, after returning the dropped term to this action \eqref{eq:GN:action2}.
Assuming that the auxiliary field is constant (corresponding to considering only the homogeneous chiral condensate), $\tilde{\sigma}(x) = \tilde{\sigma}$, we obtain the effective potential in the leading order of the $1/N$ expansion,
\begin{equation}
 V_D(\tilde{\sigma}) = \frac{\tilde{\sigma}^2 - 2m_0 \tilde{\sigma}}{2\lambda_0} - \frac{C_D}{D} (\tilde{\sigma}^2)^{D/2},
  \label{eq:effective-potential}
\end{equation}
with a constant, $C_D = \tr\!I\, (4\pi)^{-D/2} \Gammaf{1-D/2}$.

In the massless case, $m_0=0$, it is well-known fact that, under a certain renormalization condition, the gap equation, $\left.\partial V_D(\tilde{\sigma}) / \partial \tilde{\sigma} \right|_{\tilde{\sigma}=m_\chi}= 0$, can be solved exactly.
For instance, under the renormalization condition,
\begin{equation}
 \left.\frac{\partial^2 V_D(\tilde{\sigma})}{\partial \tilde{\sigma}^2}\right|_{\tilde{\sigma}=\mu_r} = \frac{\mu_r^{D-2}}{\lambda_r},
  \label{eq:renormalization:coupling}
\end{equation}
where $\mu_r$ and $\lambda_r$ stand for a renormalization scale and a renormalized coupling respectively, the solution is given by,
\begin{equation}
 m_\chi = \left[\frac{C_D^{-1}}{\lambda_r} + D-1\right]^{\frac{1}{D-2}} \mu_r.
  \label{eq:gap-equation:massless:solution}
\end{equation}
This solution \eqref{eq:gap-equation:massless:solution} is positive real when the coupling is larger than the critical one, $\lambda_\chi = C_D^{-1}(1-D)^{-1}$.
In two dimensions $\lambda_\chi = 0$ and in otherwise $\lambda_\chi > 0$.
For $\lambda_r > \lambda_\chi$, the chiral symmetry becomes spontaneously broken and the fermion mass is dynamically generated.
The coupling larger (smaller) than $\lambda_\chi$ is called the strong (weak) coupling.

The effective potential with the renormalized coupling still contains a divergence related to the bare mass parameter.
To remove the divergence, we also have to renormalize the mass parameter.
We here choose a renormalization condition,
\begin{equation}
 \frac{1}{\mu^{D-1}} \left.\frac{\partial V_D(\tilde{\sigma})}{\partial \tilde{\sigma}} \right|_{\tilde{\sigma}=\mu_r}
  = \frac{1}{\lambda_r}\left(1 - \frac{m_r}{\mu}\right) + C_D(D-2),
\end{equation}
with $m_r$ denoting the renormalized mass parameter.
Under this condition, the mass parameter gives the tilt of the effective potential, and the solution of the gap equation converges to $m_r$ in the weak coupling limit, $\lambda_r \to 0$;
in the following, we rename $m_r$ a current mass.
Thus the renormalized effective potential is given by
\begin{equation}
 V_D(\tilde{\sigma}) = \left( \frac{1}{\lambda_r} + C_D (D-1) \right) \frac{\tilde{\sigma}^2 \mu_r^{D-2} }{2} - \frac{m_r \tilde{\sigma} \mu_r^{D-2}}{\lambda_r} - \frac{C_D}{D} \left(\tilde{\sigma}^2\right)^{D/2}.
  \label{eq:effective-potential:massive}
\end{equation}
In the two-dimensional limit, it is confirmed that this expression is consistent with Ref.~\cite{Barducci:1994cb}.

To consider the GN model at a finite temperature, $T$, and a chemical potential, $\mu$, we modify the action \eqref{eq:GN:action1} by using the Matsubara formalism.
The effective potential reads
\begin{equation}
 \begin{aligned}
  V_D(\tilde{\sigma}; m_r, \mu_r; T, \mu)
  = &\left( \frac{1}{\lambda_r} + (D-1)C_D\right) \frac{\tilde{\sigma}^2 \mu_r^{D-2}}{2} - \frac{m_r \tilde{\sigma} \mu_r^{D-2}}{\lambda_r} - \frac{C_D}{D} (\tilde{\sigma}^2)^{D/2} \\
  &-  \tilde{C}_D T \int^\infty_0 \dmeasure{}{q} q^{D-2} \left[ \ln \left(1+e^{-\left(\sqrt{q^2 +\tilde{\sigma}^2} -\mu \right)/T}\right) + (-\mu \to \mu )\right],
 \end{aligned}
 \label{eq:effective-potential:massive:temp-chem}
\end{equation}
where $\tilde{C}_D = (4\pi)^{-(D-1)/2} \tr I /\Gammaf{\frac{D-1}{2}}$.
In this expression, the thermal part (the second line in Eq. \eqref{eq:effective-potential:massive:temp-chem}) is separated from the vacuum part (the first line in Eq. \eqref{eq:effective-potential:massive:temp-chem}).
We use the symbol $M$ as the solution of the gap equation, $\partial V_D(\tilde{\sigma}; m_r, \mu_r; T, \mu)/ \partial \tilde{\sigma} |_{\tilde{\sigma}=M} = 0$, in this paper;
in our definition of $\tilde{\sigma}$, the solution of the gap equation is equivalent to the dynamically generated fermion mass.

\section{Super restoration}\label{sec:super-restoration}
\subsection{Analytical results}\label{sec:analyt-results}
At zero temperature and zero chemical potential, the dynamically generated fermion mass is always not zero in the massive theory, and converges to $m_r$ from above in the weak coupling limit, $\lambda_r \to 0$.
It is also well known that the chiral symmetry tends to be restored at high temperature or large chemical potential.
In the massless theory, we can observe the first-order phase transition boundary at low temperature and large chemical potential, and the second-order one at high temperature and small chemical potential on a $\mu$-$T$ plane.
In this section, we show that the dynamical mass decreases below the current mass, $m_r$, at a high/large but finite temperature and chemical potential; we call such a phenomenon super restoration.

We consider a solution of the gap equation satisfying $M \leq m_r$ on a $\mu$-$T$ plane,
\begin{equation}
 \begin{aligned}
  0 = &\left(\frac{1}{\lambda_r} + C_D (D-1)\right) \frac{M}{\mu_r} - \frac{m_r}{\lambda_r \mu_r} - C_D \left(\frac{M}{\mu_r}\right)^{D-1} \\
  &- \tilde{C}_D \frac{M}{\mu_r}\int^\infty_0 \frac{\dmeasure{}{q}}{\sqrt{q^2+M^2}} \left(\frac{q}{\mu_r}\right)^{D-2} \left[ \left(e^{\left(\sqrt{q^2 +M^2} +\mu \right)/T} + 1\right)^{-1} + (\mu \to -\mu )\right].
 \end{aligned}
 \label{eq:gap-equation:mass-chem-temp}
\end{equation}
If such a solution exists and is continuous outer the first-order phase transition boundary on the $\mu$-$T$ plane, at least there is the solution that satisfies $M = m_r$.
Supposing $M=m_r(>0)$ in this equation, we obtain
\begin{equation}
 \begin{aligned}
  &\Gammaf{1-\frac{D}{2}}\left( D-1 - \left(\frac{m_r}{\mu_r}\right)^{D-2}\right) \\
  =& - \frac{2 \pi^{1/2}}{\Gammaf{\frac{D-1}{2}}} \int^\infty_0 \frac{\dmeasure{}{q}}{\sqrt{q^2+m_r^2}} \left(\frac{q}{\mu_r}\right)^{D-2} \left[ \left(e^{\left(\sqrt{q^2 +m_r^2} +\mu \right)/T} + 1\right)^{-1} + (\mu \to -\mu )\right].
 \end{aligned}
 \label{eq:gap-equation:mr-div}
\end{equation}

This expression can be taken the chiral limit, $M=m_r \to 0$.
The reduced expression is given by
\begin{equation}
 \frac{D-1}{2 \pi^{1/2}}\Gammaf{1-\frac{D}{2}}\Gammaf{\frac{D-1}{2}}
  = \Gammaf{D-2} \left(\frac{T}{\mu_r}\right)^{D-2} \left[ \PolyLog{D-2}{-e^{-\mu/T}} + \PolyLog{D-2}{-e^{\mu/T}} \right],
  \label{eq:gap-equation:mr-div:chiral}
\end{equation}
where $\PolyLog{a}{s}$ is a polylogarithm.
Equations \eqref{eq:gap-equation:mr-div} and \eqref{eq:gap-equation:mr-div:chiral} does not indicate the existence of the solution, but give the way to confirm it.
At zero chemical potential, the equation can be solved exactly as
\begin{equation}
 \frac{\tilde{T}}{\mu_r}
  = \left( -\frac{2^{D-1}-4}{D-1} \frac{\Gammaf{\frac{D}{2}-1}}{\Gammaf{1-\frac{D}{2}}} \zetaf{D-2}\right)^{\frac{1}{2-D}},
  \label{eq:gap-equation:mr-div:chiral:zero-temp}
\end{equation}
with $\tilde{T}$ denoting the temperature at which the dynamical mass becomes $m_r$ in the chiral limit.
The specific values are $e^{1+\gamma}/\pi \simeq 1.54$ in $D=2$ and $1/\ln 2 \simeq 1.44$ in $D=3$.
They are about two to three times higher than the critical temperature ($e^\gamma/\pi \simeq 0.57$ in $D=2$ and $1/\ln 4 \simeq 0.72$ in $D=3$) in the massless theory \cite{Wolff:1985av, Inagaki:1994ec}.
In the same way, we can solve exactly the equation at zero temperature,
\begin{equation}
 \frac{\tilde{\mu}}{\mu_r}
  = \left( -\frac{(D-1)(D-2)}{2\pi^{1/2}}\Gammaf{1-\frac{D}{2}}\Gammaf{\frac{D-1}{2}}\right)^{\frac{1}{D-2}},
  \label{eq:gap-equation:mr-div:chiral:zero-chem}
\end{equation}
with $\tilde{\mu}$ denoting the chemical potential at which the dynamical mass becomes $m_r$ in the chiral limit.
The specific values are $e/2 \simeq 1.36$ in $D=2$ and $2$ in $D=3$.
These are also about two times larger than the critical chemical potential ($1/\sqrt{2} \simeq 0.71$ in $D=2$ and $1$ in $D=3$) in the massless theory \cite{Wolff:1985av, Inagaki:1994ec}.
The super restoration boundary in the chiral limit is described by Eq. \eqref{eq:gap-equation:mr-div:chiral} as the curve connecting $\tilde{T}$ and $\tilde{\mu}$ on the $\mu$-$T$ plane.
For a current mass enough small, the super restoration boundary in the chiral limit gives approximately the boundary at a finite current mass.

\subsection{Numerical results}\label{sec:numer-results}
The solution of the gap equation in the chiral limit, $M=m_r \to 0$, has been found analytically; in particular, we have found the specific values at zero temperature or chemical potential.
We here calculate numerically the super restoration boundary at a finite $m_r$ based on Eq. \eqref{eq:gap-equation:mass-chem-temp} and plot it on the $\mu$-$T$ plane.
In our calculations, we set the coupling constant satisfying $M/\mu_r=1$ in the trivial condition ($m_r=0$, $T=0$ and $\mu=0$) for arbitrary dimensions and $\tr\!I = 2^{D/2}$.

\begin{figure}[htb]
 \centering
 \begin{minipage}{0.85\hsize}
  \centering
  \begin{tabular}{cc}
   \begin{minipage}{0.4\hsize}
    \centering
    \includegraphics[width=1\hsize,clip]
    {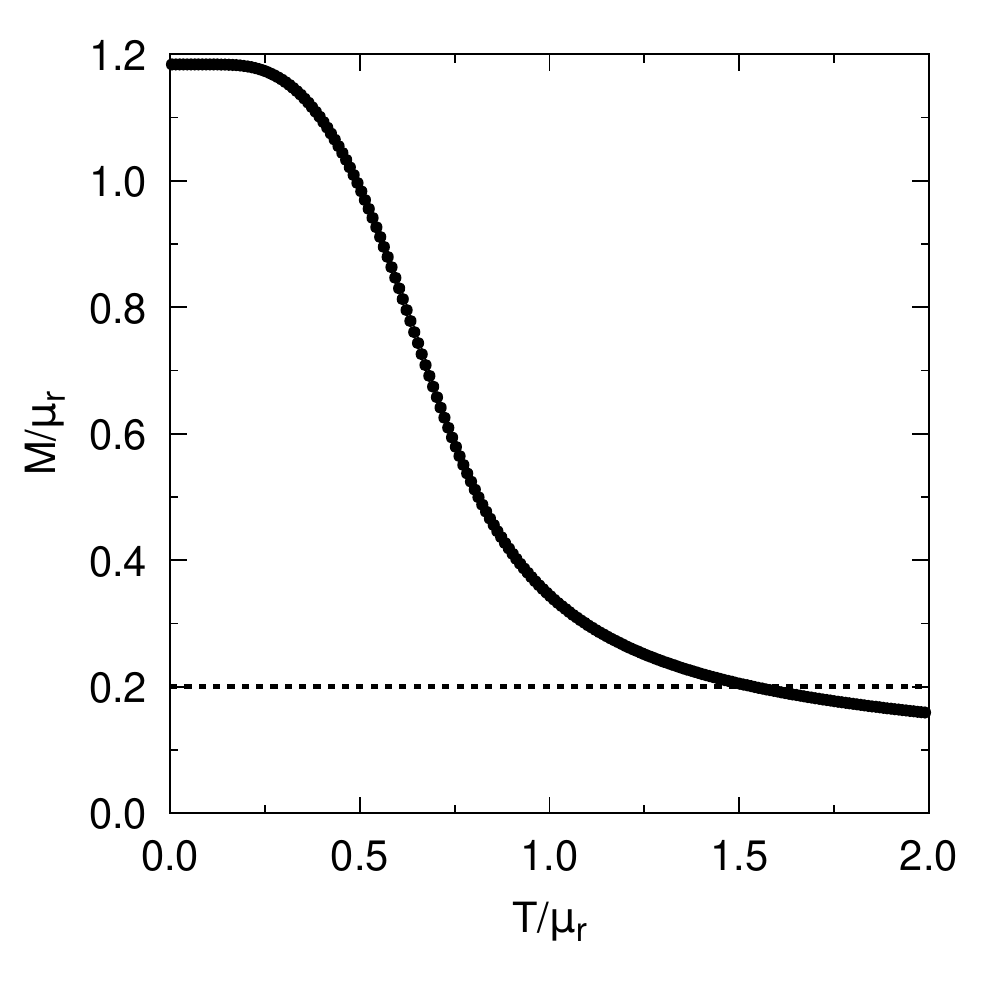}
   \end{minipage}
   &
   \begin{minipage}{0.4\hsize}
    \centering
    \includegraphics[width=1\hsize,clip]
    {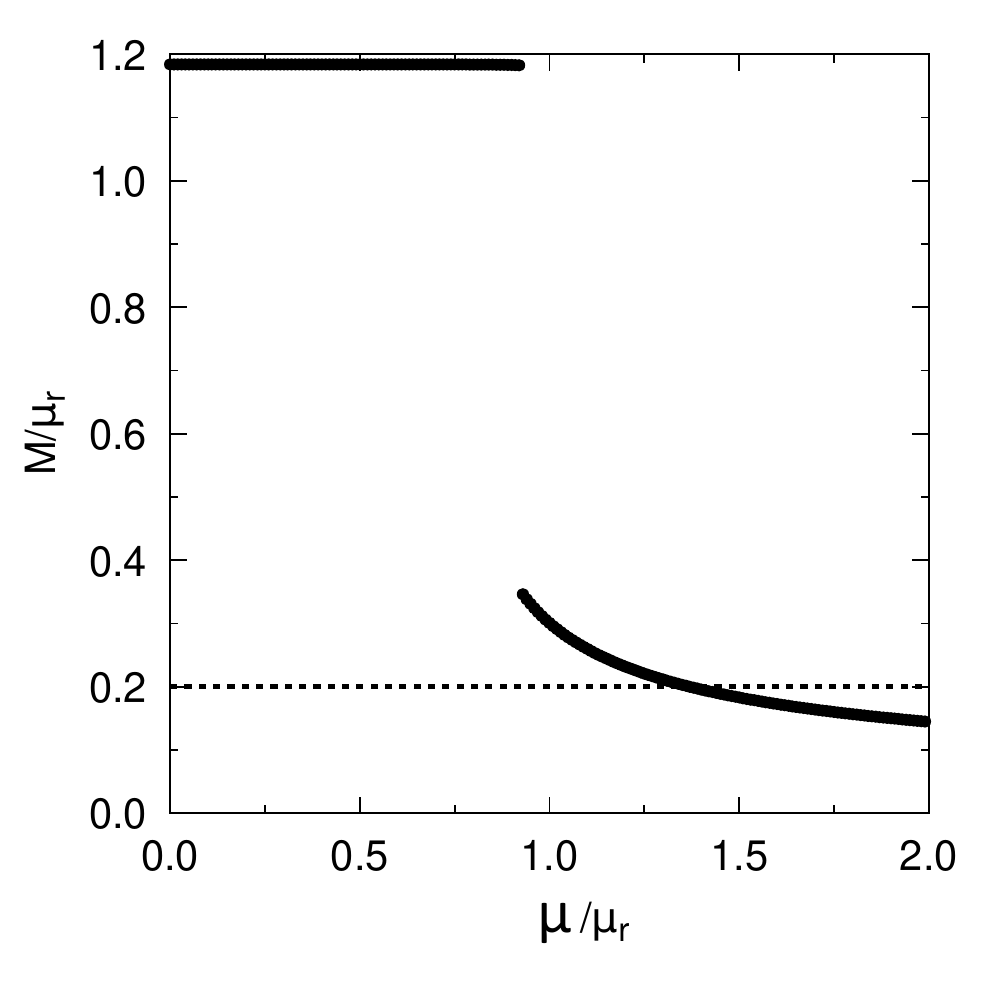}
   \end{minipage}
  \end{tabular}
  \caption{Behavior of the dynamically generated fermion mass as a function of temperature (left, fixed $\mu/\mu_r=0.05$) and chemical potential (right, fixed $T/\mu_r=0.05$) in $D=2$.
  The current mass is shown by a dotted line ($m_r=0.2$).}
  \label{fig:dynamical-mass:2dim}
 \end{minipage}
\end{figure}

\begin{figure}[htb]
 \centering
 \begin{minipage}{0.85\hsize}
  \centering
  \begin{tabular}{cc}
   \begin{minipage}{0.4\hsize}
    \centering
    \includegraphics[width=1\hsize,clip]
    {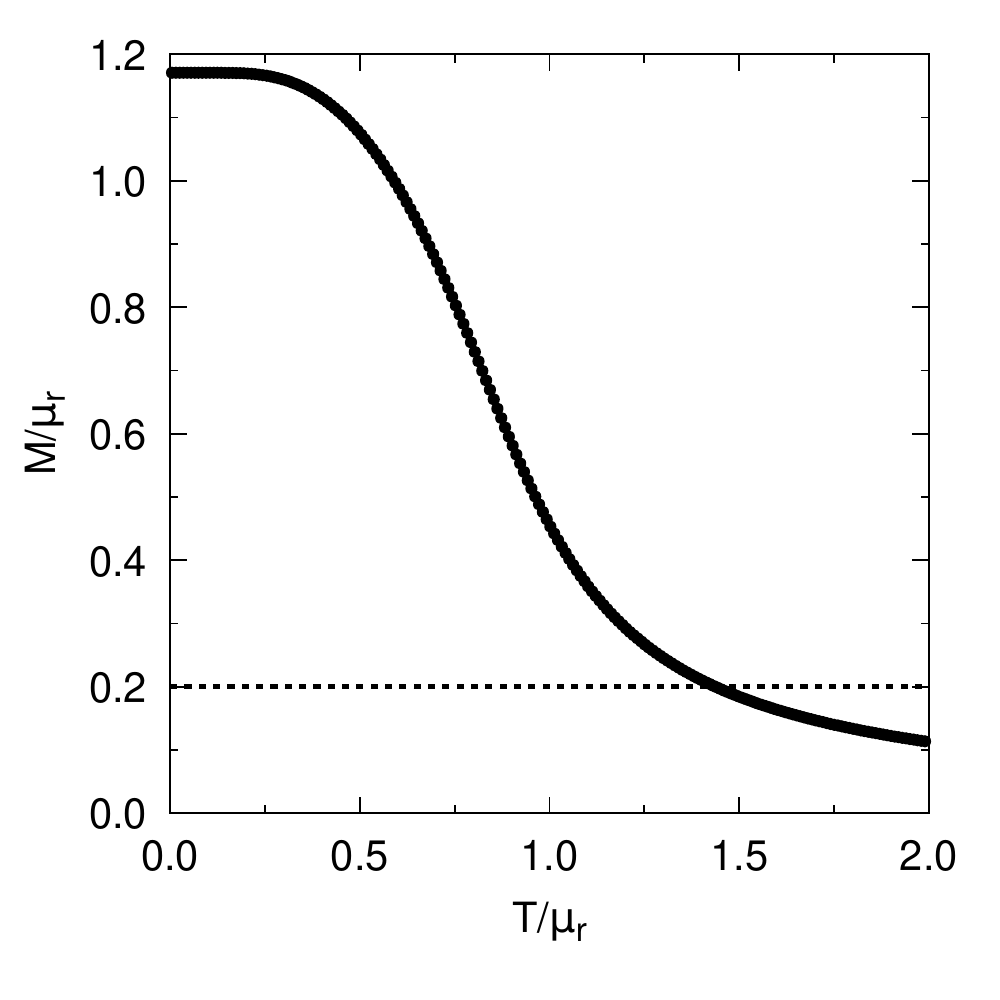}
   \end{minipage}
   &
   \begin{minipage}{0.4\hsize}
    \centering
    \includegraphics[width=1\hsize,clip]
    {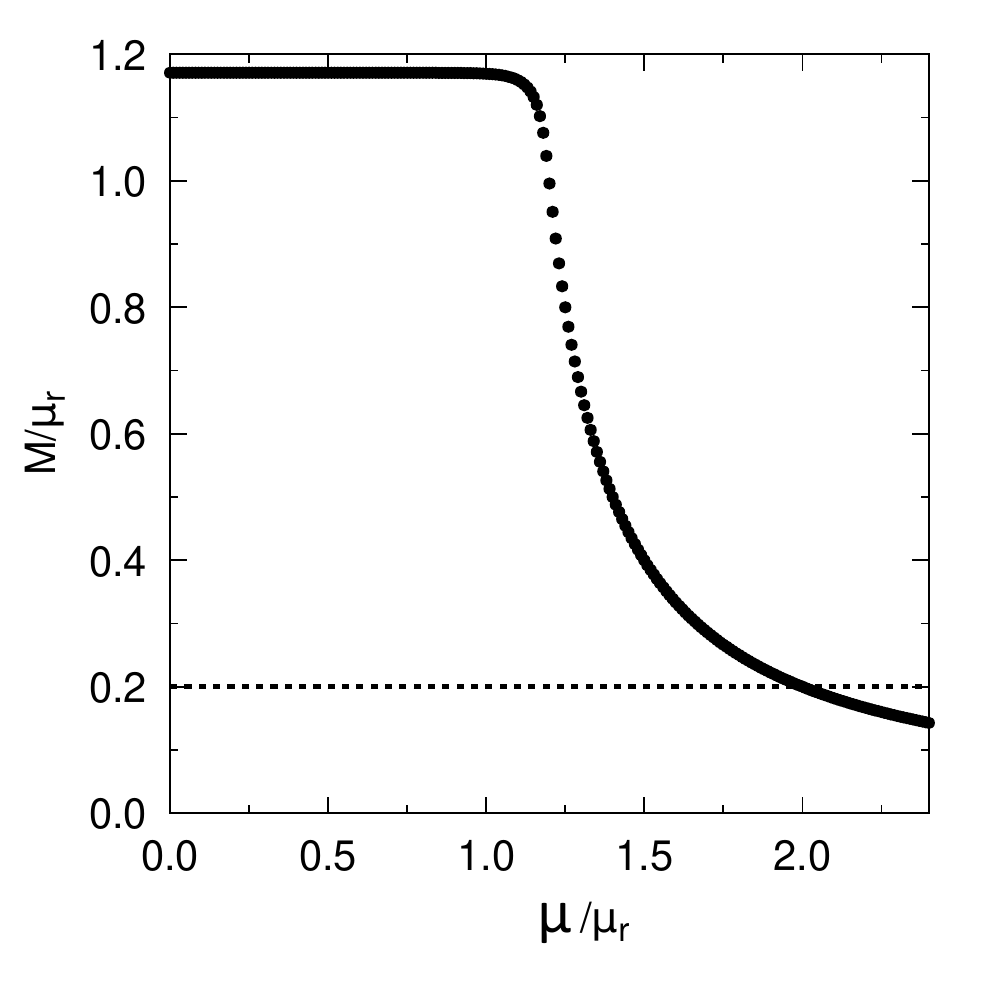}
   \end{minipage}
  \end{tabular}
  \caption{Behavior of the dynamically generated fermion mass as a function of temperature (left, fixed $\mu/\mu_r=0.05$) and chemical potential (right, fixed $T/\mu_r=0.05$) in $D=3$.
  The current mass is shown by a dotted line ($m_r=0.2$).}
  \label{fig:dynamical-mass:3dim}
 \end{minipage}
\end{figure}

Behavior of the dynamically generated fermion mass as a function of temperature and chemical potential is shown in Figs. \ref{fig:dynamical-mass:2dim} ($D=2$) and \ref{fig:dynamical-mass:3dim} ($D=3$) with a fixed current mass, $m_r=0.2$.
On the left figure, a second-order phase transition is replaced with a crossover due to the massive theory.
As the figures show, the dynamical fermion mass smoothly becomes below the current mass at high temperature or large chemical potential.

\begin{figure}[htb]
 \centering
 \begin{minipage}{0.85\hsize}
  \centering
  \begin{tabular}{cc}
   \begin{minipage}{0.48\hsize}
    \centering
    \includegraphics[width=1\hsize,clip]
    {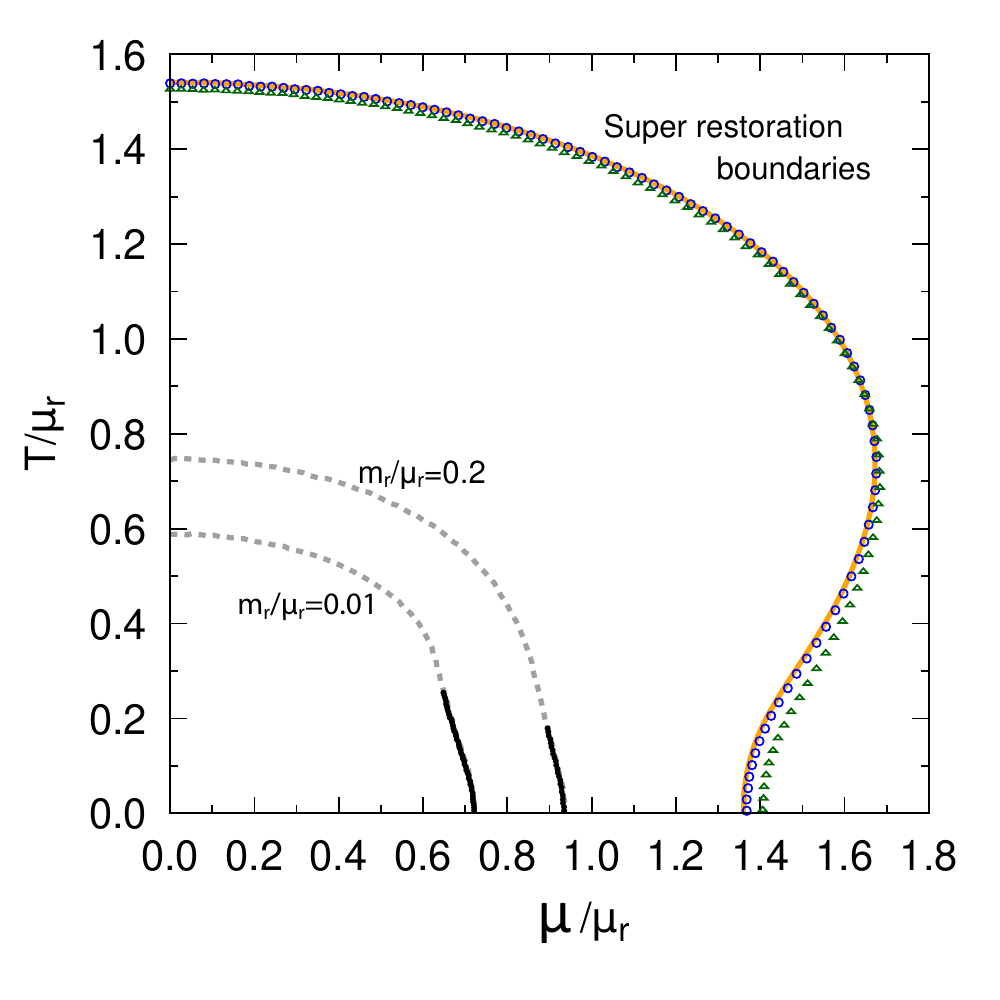}
    \subcaption{$D=2$}
   \end{minipage}
   &
   \begin{minipage}{0.48\hsize}
    \centering
    \includegraphics[width=1\hsize,clip]
    {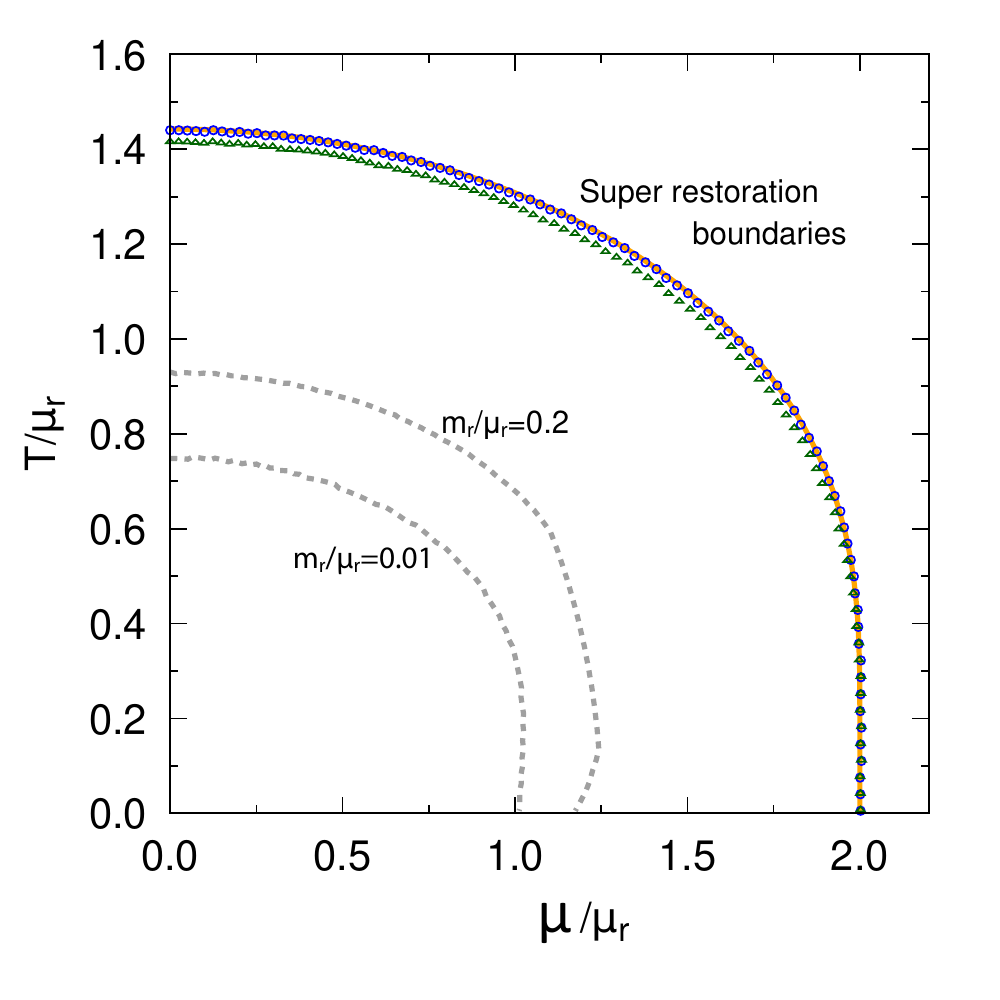}
    \subcaption{$D=3$}
   \end{minipage}
  \end{tabular}
  \caption{Behavior of (outer) the super restoration boundaries (solid: the chiral limit, circles: $m_r=0.2$, and triangles: $m_r=0.5$) with (inner: $m_r=0.01$ and middle: $m_r=0.2$ lines) the first-order phase transition boundaries (solid) and the peaks of the chiral susceptibility (dotted).}
  \label{fig:boundaries}
 \end{minipage}
\end{figure}

The super restoration boundaries are shown as outer lines in Fig. \ref{fig:boundaries};
the solid and dotted lines denote the chiral limit and the massive cases with $m_r/\mu_r=0.2$ (circles) and $0.5$ (triangles), respectively.
To compare with these, we also plot first-order phase transition boundaries (solid) and peaks of the chiral susceptibility (dotted) that is defined by $\partial M/ \partial m_r$ at $m_r/\mu_r=0.01$ (inner) and $m_r/\mu_r=0.2$ (middle).
In three dimensions, there is no phase boundary and the lines of the chiral susceptibility are curved to extend the area enclosed by the lines at low temperature and large chemical potential.
Since the lines are drawn in terms of the susceptibility, it does not necessarily coincide with the lines where the dynamical mass changes the most.

The figures show that the super restoration boundaries almost overlap each other.
The area enclosed by them tends to shrink in particular at high temperature and small chemical potential.
On the other hand, the area tends to expand at low temperature and large chemical potential in $D=2$.
These indicate the different behavior of the super restoration boundaries from the first-order phase transition boundary and the peaks which expand monotonically with increasing $m_r$.
It is also observed that the difference of the super restoration boundaries caused by the current mass change is slighter than the boundary of the first-order phase transition and the peaks of the chiral susceptibility.

\section{Two-flavor NJL model}\label{sec:two-flavor-njl}
In this section, we apply the previous discussion to the two-flavor NJL model on the four-dimensional spacetime.
The Lagrangian of the two-flavor NJL model with current quark mass is given as,
\begin{equation}
 \mathcal{L} = \psibar(i \gamma^\mu \partial_\mu - m)\psi
  + \frac{G}{2N_c} \left[ (\psibar\psi) ^2+ (\psibar i\gamma_5 \tau^a  \psi)^2 \right],
\end{equation}
with a mass matrix $m=\mathrm{diag}(m_u, m_d)$, an effective coupling constant $G$, the number of colors $N_c$ and the Pauli matrices of the isospin vector $\tau^a (a=1,2,3)$.
For simplicity, we assume $m_u = m_d$.
While the Lagrangian has the $SU(2)_L \times SU(2)_R$ global symmetry at the massless limit, the symmetry is explicitly broken down to the $SU(2)_{L+R}$ because of the non-zero quark masses.
Since the effective coupling has the negative mass dimensions and the model cannot be renormalizable, one has to introduce the momentum cutoff to evaluate the physical quantities.

In the leading order of the $1/N_c$ expansion, we obtain the effective potential,
\begin{equation}
 V(\sigma) = \frac{\sigma^2}{2G} - \frac{1}{2N_c} \int \frac{\dmeasure{4}{q}}{i(2\pi)^4} \ln \det(\gamma^\mu q_\mu - m - \sigma + i\epsilon),
\end{equation}
with the auxiliary scalar field $\sigma \simeq -(G/N_c) \psibar \psi$.
From the stationary condition of the effective potential, the gap equation is expressed as follows,
\begin{equation}
 \Braket{\sigma} = 2G \cdot i \tr S(M),
\end{equation}
where $M$ is the constituent mass $M = m_u + \Braket{\sigma}$, the trace is the sum over spinor indices and $S(M)$ is the quark propagator,
\begin{equation}
 i\tr S(M) = - \tr \int \frac{\dmeasure{4}{q}}{i(2\pi)^4} \frac{1}{\gamma^\mu q_\mu - M +i\epsilon}.
\label{itrS}
\end{equation}
The momentum integral in Eq.~\eqref{itrS} is divergent, and we introduce the momentum cutoff scale later.

Since we are interested in the behavior of the constituent mass in a thermal system, we apply the imaginary time formalism with a chemical potential to the model.
By using the formalism, the gap equation at a finite temperature and chemical potential is derived as follows,
\begin{align}
 \Braket{\sigma} &= 2G \left[ i\tr S^0(M, \Lambda) + i\tr S^T(M, \Lambda) \right],
 \label{gap_t} \\
 i\tr S^0(M, \Lambda) &= \frac{M}{\pi^2} \int_0^\Lambda \frac{\dmeasure{}{q} q^2}{E(M)},
 \label{itrS0} \\
 i\tr S^T(M, \Lambda) &= -\frac{M}{\pi^2} \int_0^\Lambda \frac{\dmeasure{}{q} q^2}{E(M)} \left(\frac1{1+e^{E^+/T}} + \frac1{1+e^{E^-/T}} \right),
 \label{itrSt}
\end{align}
with $E(M)=\sqrt{q^2+M^2}$ and $E^\pm = E(M)\pm\mu$.
Equation~\eqref{itrS0} is the zero temperature part of the gap equation and is equal to Eq.~\eqref{itrS} integrated with respect to $q_0$.
We regularize the integral in Eq.~\eqref{itrS0} using three-momentum cutoff scale $\Lambda$, and adopt the values of Ref.~\cite{Hatsuda:1994pi}, $\Lambda=\SI{631}{MeV}$ and $G/(2N_c)\simeq \SI{5.51E-6}{MeV^{-2}}$.

Equation~\eqref{itrSt} is the finite temperature and chemical potential part of the gap equation.
Unlike the zero temperature part, the momentum integral is finite for $\Lambda \to \infty$.
In this paper,  we consider two regularizations in the integral of Eq.~\eqref{itrSt} because the value is finite whether the limit is taken or not.
In the case 1, the same cutoff scale is used as the zero temperature part.
In the case 2, the momentum scale is integrated out and Eq.~\eqref{gap_t} is written as follows,
\begin{equation}
 \Braket{\sigma} = 2G \left[ i\tr S^0(M, \Lambda) + i\tr S^T(M, \infty) \right].
\end{equation}
In this case, the effect of the finite temperature and chemical potential is independent of the regularization scale.

The effective potential at a finite $T$ and $\mu$ is written as follows,
\begin{align}
 V(\sigma) &= \frac{\sigma^2}{2G}+V^0(\sigma,\Lambda)+V^T(\sigma,\Lambda) ,\\
 V^0(\sigma,\Lambda) &= -\frac{2}{\pi^2} \int_0^\Lambda \dmeasure{}{q} q^2E(M) ,\\
 V^T(\sigma,\Lambda) &= -\frac{2T}{\pi^2} \int_0^\Lambda \dmeasure{}{q} q^2\left[ \ln\left(1+e^{-E^+/T} \right) + \ln\left(1+e^{- E^-/T}\right)\right].
\end{align}
In the case 2, we use the following effective potential,
\begin{equation}
 V(\sigma) = \frac{\sigma^2}{2G}+V^0(\sigma,\Lambda)+V^T(\sigma,\infty) .
\end{equation}

\begin{figure}[htb]
 \centering
 \begin{minipage}{0.85\hsize}
  \centering
  \begin{tabular}{cc}
   \begin{minipage}{0.48\hsize}
    \centering
    \includegraphics[width=1\hsize,clip]
    {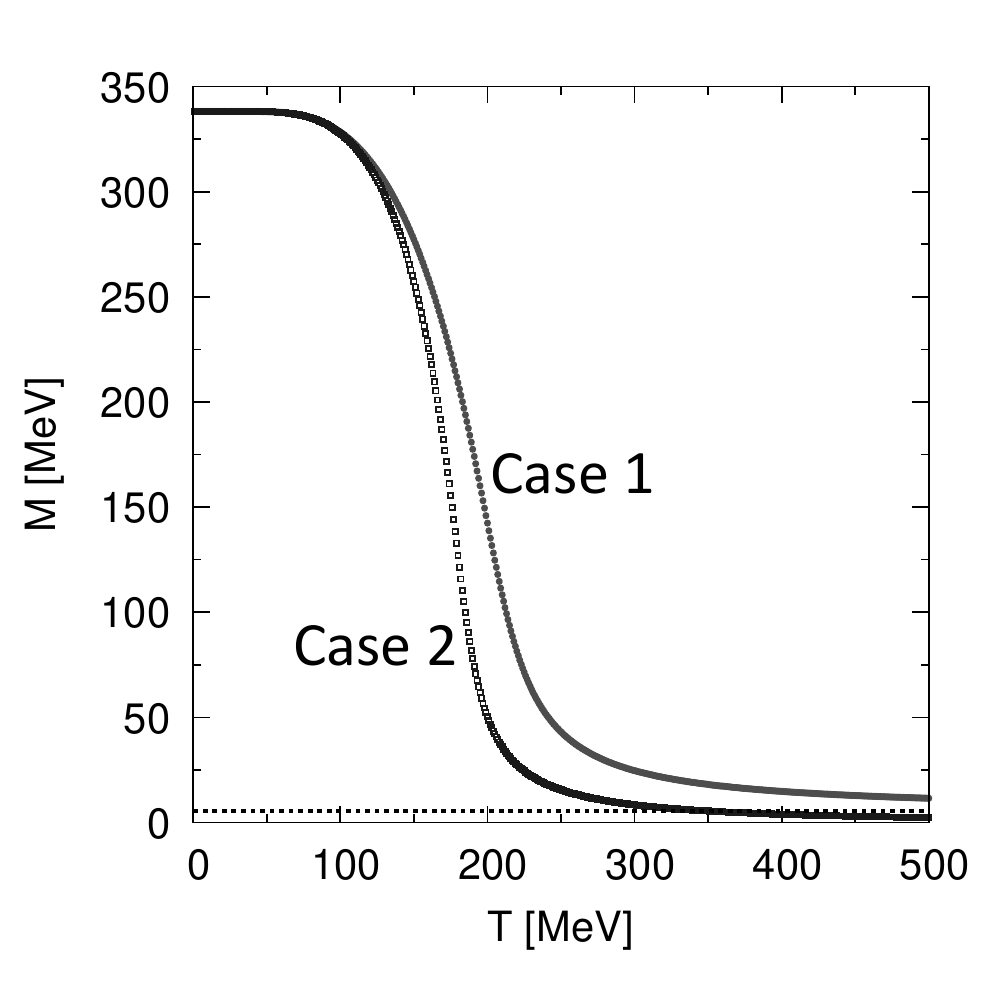}
   \end{minipage}
   &
   \begin{minipage}{0.48\hsize}
    \centering
    \includegraphics[width=1\hsize,clip]
    {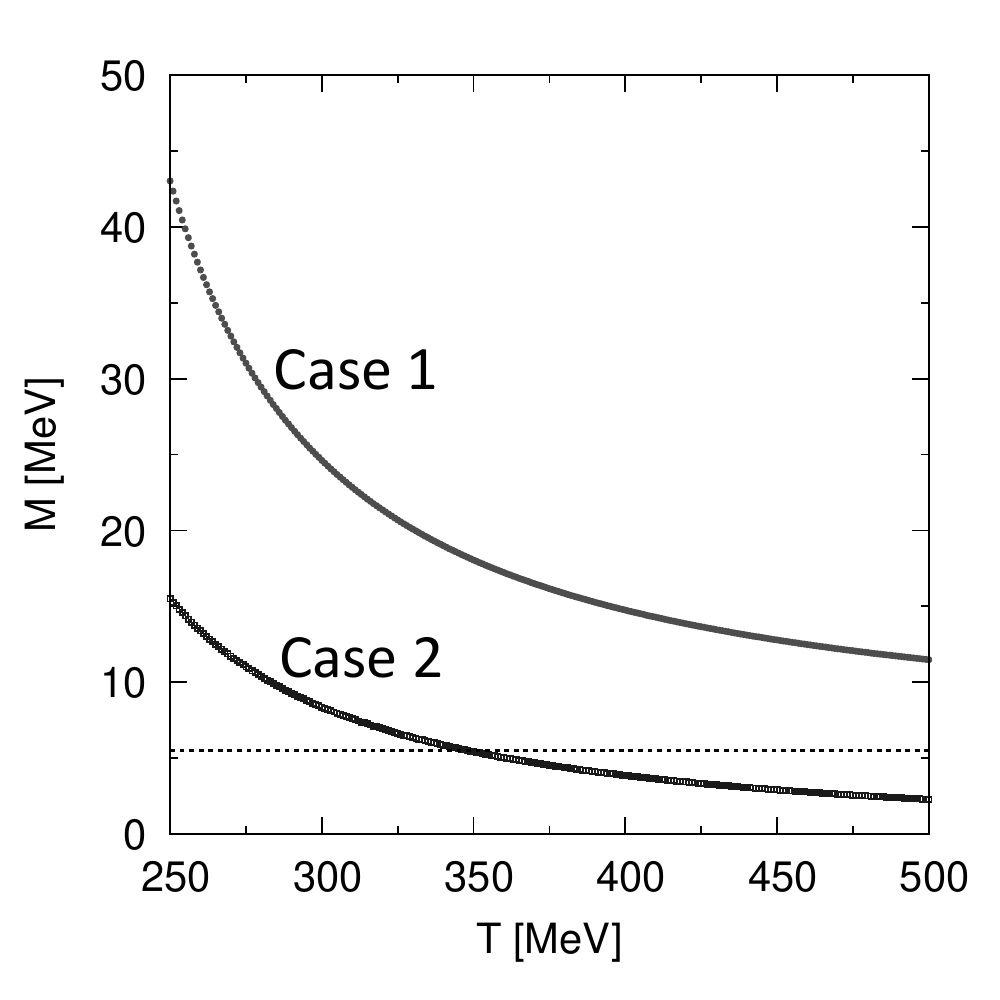}
   \end{minipage}
  \end{tabular}
  \caption{Constituent mass $M$ as a function of $T$ at $\mu=0$ in the two cases.
  Horizontal dotted lines represent the value of the current quark mass.
  Right figure shows the high temperature region of the behavior for the constituent mass.}
  \label{fig:gap-t}
 \end{minipage}
\end{figure}

We show the behavior of the constituent mass as a function of $T$ in Fig.~\ref{fig:gap-t}.
The thermal effect reduces the dynamical mass.
As seen in the left figure, the thermal effect in the case 2 is larger than that in the case 1.
At high temperature, it is found that the constituent mass, $M$, becomes smaller than the current quark mass, $m_u = \SI{5.5}{MeV}$,  in the case 2.
The super restoration takes place in the case 2, but not in the case 1.

\begin{figure}[htb]
 \centering
 \begin{minipage}{0.85\hsize}
  \centering
  \begin{tabular}{cc}
   \begin{minipage}{0.48\hsize}
    \centering
    \includegraphics[width=1\hsize,clip]
    {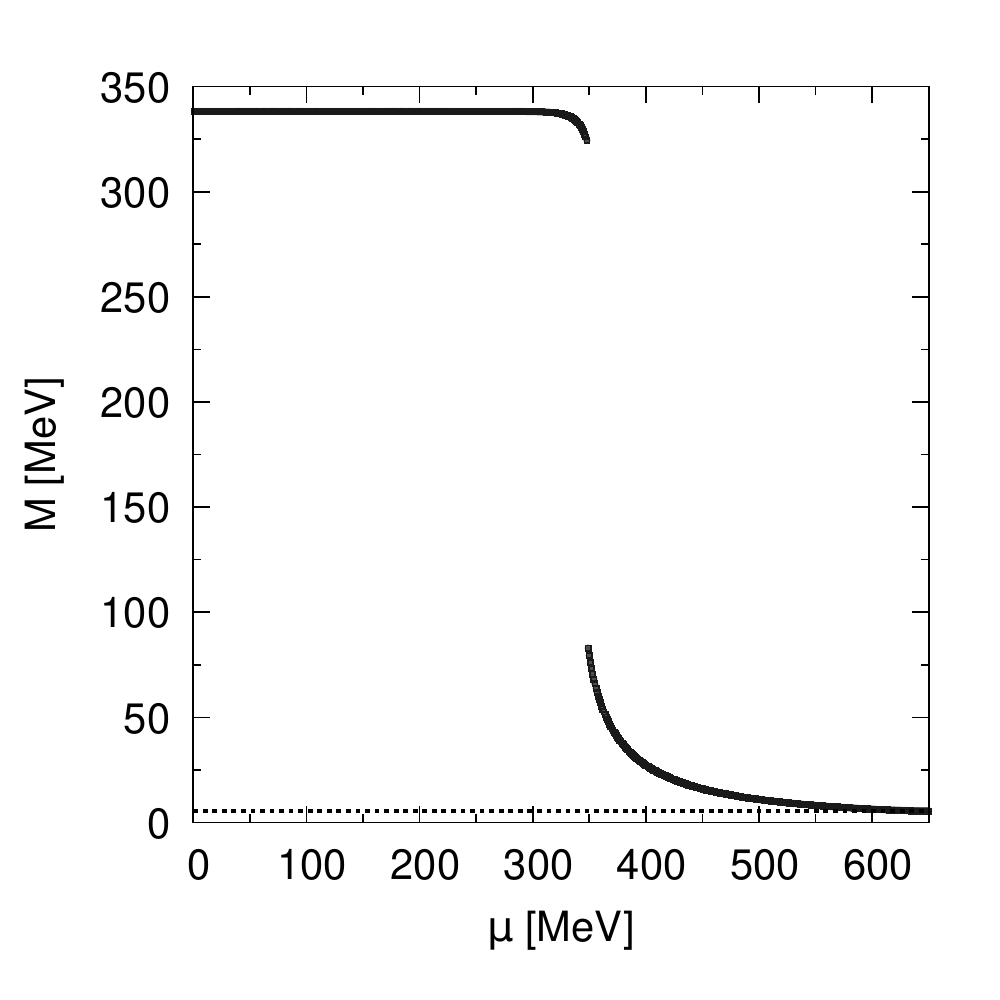}
   \end{minipage}
   &
   \begin{minipage}{0.48\hsize}
    \centering
    \includegraphics[width=1\hsize,clip]
    {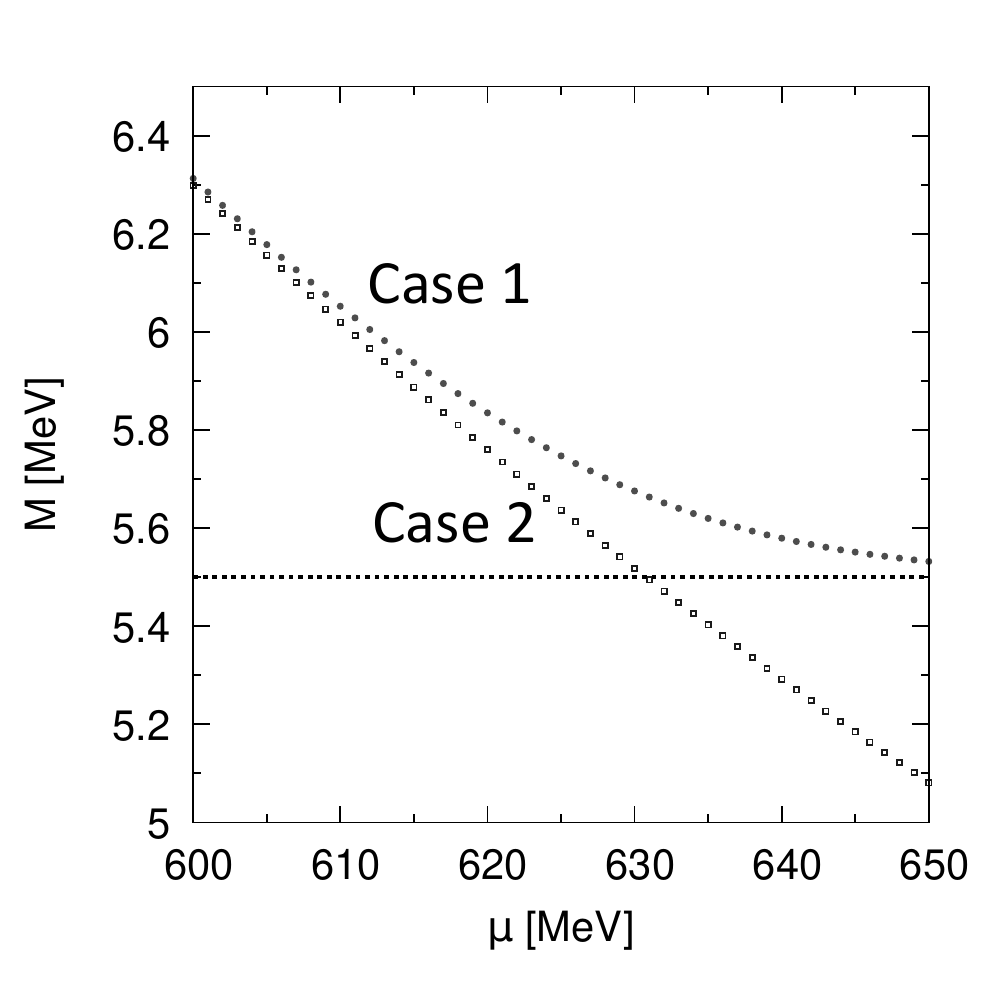}
   \end{minipage}
  \end{tabular}
  \caption{Constituent mass $M$ as a function of $\mu$ at $T=\SI{10}{MeV}$ in the two cases.
  Horizontal dotted lines represent the value of the current quark mass.
  Right figure shows the large chemical potential region of the behavior for the constituent mass.}
  \label{fig:gap-mu}
 \end{minipage}
\end{figure}

We show the behavior of the constituent mass as a function of $\mu$ at $T=\SI{10}{MeV}$ in Fig.~\ref{fig:gap-mu}.
From the left figure, the effects of the chemical potential are indistinguishable in each case.
The right figure shows that the constituent mass, $M$, approaches the current quark mass, $m_u$, and becomes smaller than $m_u$ for large chemical potential, $\mu \gtrsim \Lambda$, in the case 1 and 2, respectively.

\begin{figure}[htb]
 \centering
 \begin{minipage}{0.85\hsize}
  \centering
  \begin{tabular}{cc}
   \begin{minipage}{0.48\hsize}
    \centering
    \includegraphics[width=1\hsize,clip]
    {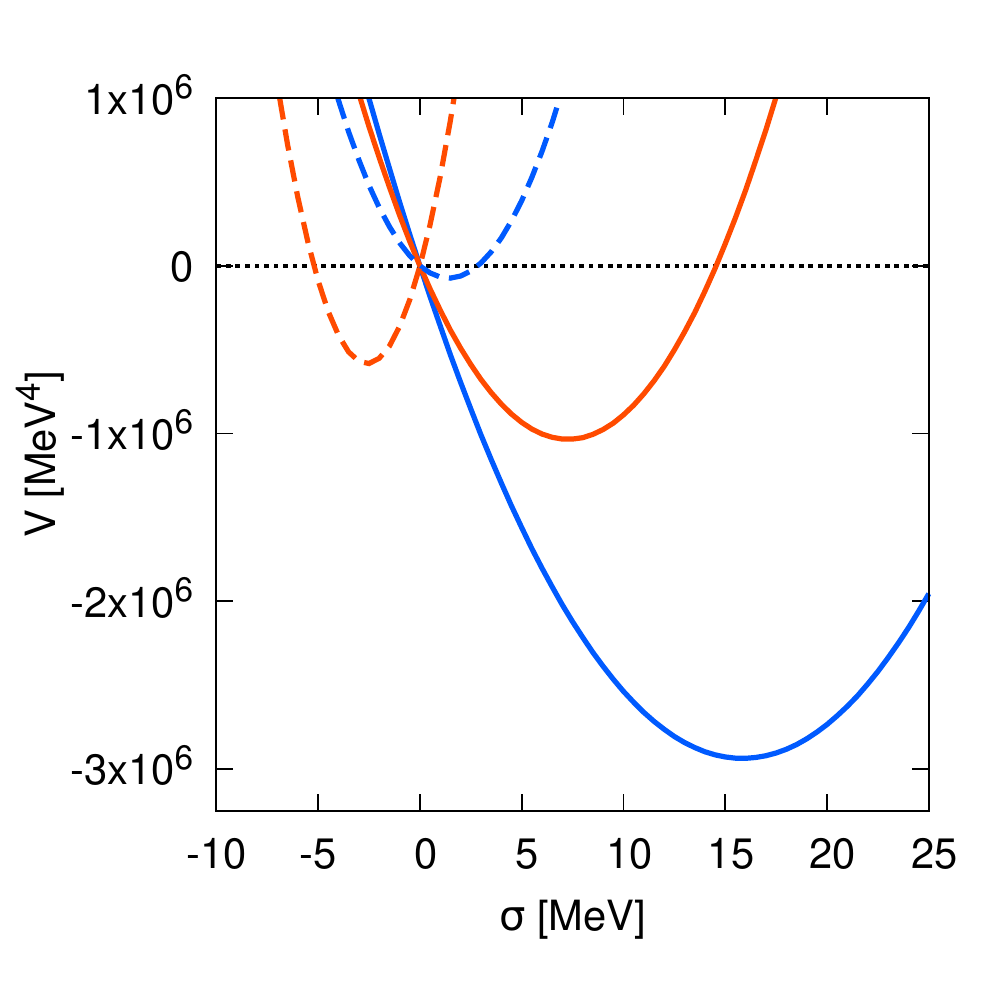}
   \end{minipage}
   &
   \begin{minipage}{0.48\hsize}
    \centering
    \includegraphics[width=1\hsize,clip]
    {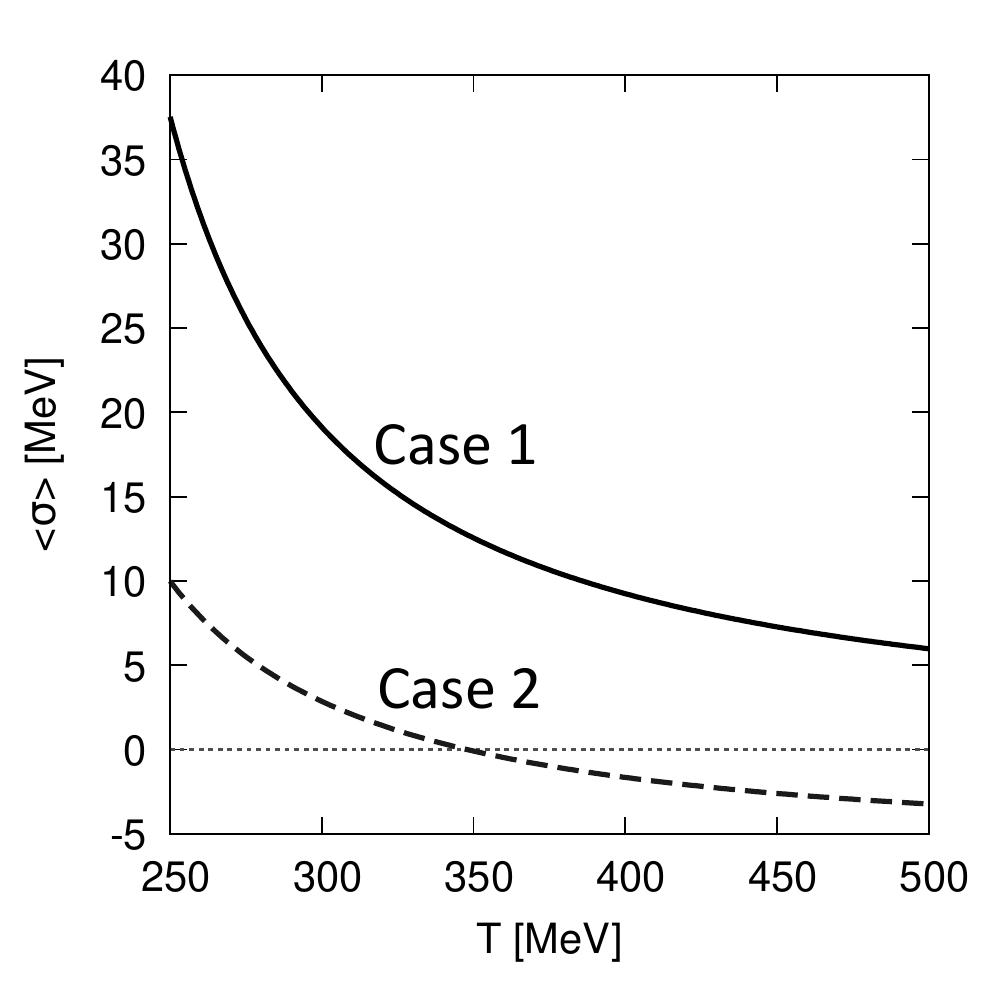}
   \end{minipage}
  \end{tabular}
  \caption{(Left) Behavior of the effective potential is shown as a function of $\sigma$ at $\mu = 0$.
  The solid and dashed lines represent the effective potential in the case 1 and 2, respectively.
  The colors of the lines display temperatures (blue: $T=\SI{320}{MeV}$, and red: $\SI{450}{MeV}$).
  (Right) The minimum of the effective potential as a function of $T$ at $\mu=0$.
  The line types represent the cases as well as the left.}
  \label{fig:Veff}
 \end{minipage}
\end{figure}

We show the behavior of the effective potential in the high temperature region in Fig.~\ref{fig:Veff}.
In the case 1 (solid), the minimum of the effective potential \ie the expectation value of $\sigma$ becomes smaller as $T$ increases.
Since the chiral symmetry is explicitly broken due to the current quark mass, the minimum of the effective potential does not become 0.
On the other hand, in the case 2 (dashed), the minimum of the effective potential becomes 0 at $T \simeq \SI{350}{MeV}$.
The expectation value of $\sigma$ changes from positive to negative as $T$ increases.
The negative value of $\Braket{\sigma}$ causes the constituent quark masses to be smaller than the current quark masses.
This means that the chiral condensate takes place to counteract the explicit symmetry breaking at high temperature.

\begin{figure}[htb]
 \centering
 \begin{minipage}{0.85\hsize}
  \centering
  \begin{tabular}{c}
   \begin{minipage}{0.5\hsize}
    \centering
    \includegraphics[width=1\hsize,clip]
    {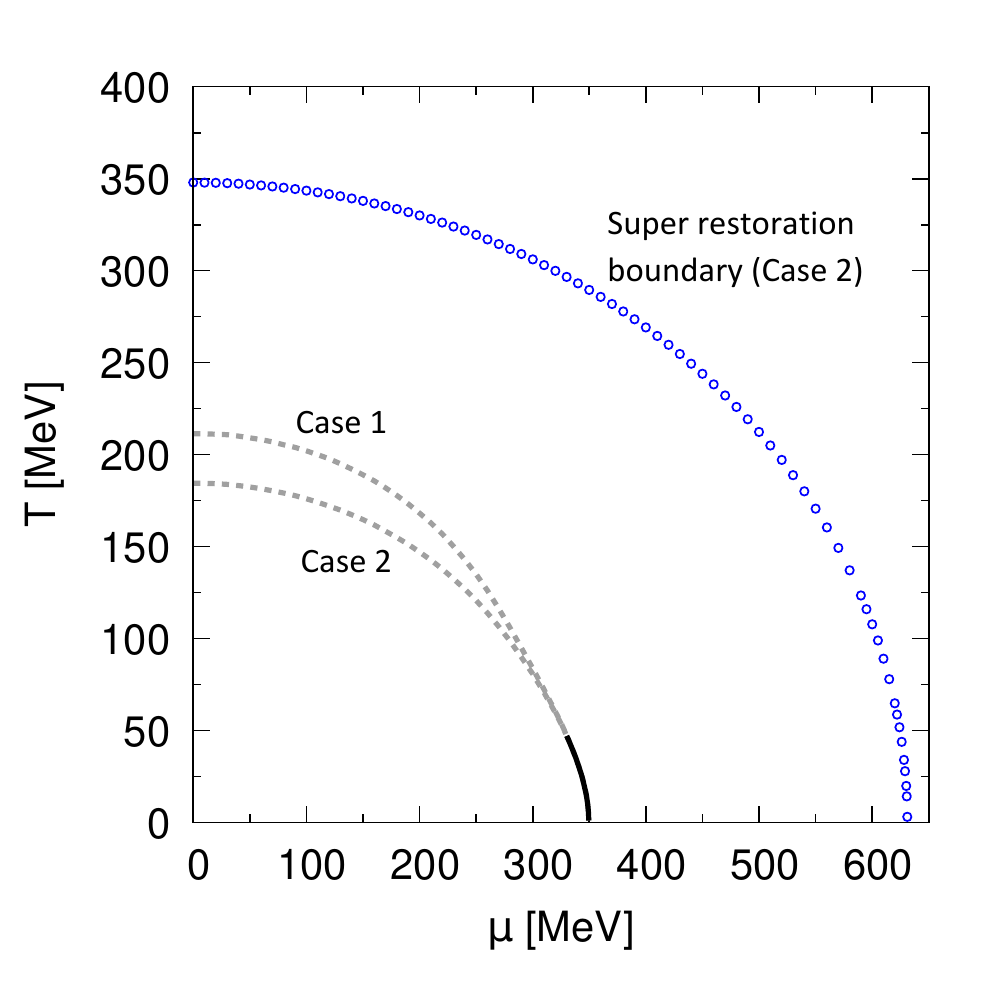}
   \end{minipage}
  \end{tabular}
  \caption{Behavior of the super restoration boundary (blue circles) with the first-order phase transition boundary (black solid) and the peaks of the chiral susceptibility (gray dotted: The outer and inner correspond case 1 and 2, respectively).}
  \label{fig:chiss}
 \end{minipage}
\end{figure}

We draw the peaks of the chiral susceptibility \cite{Fukushima:2003fw, Zhao:2006br} and the super restoration boundary in Fig.~\ref{fig:chiss}.
The area enclosed by the line of the maximum of the chiral susceptibility in the case 2 is smaller than that in the case 1.
This feature is related to the fact that the thermal contribution in the case 2 is larger than the case 1 in Fig.~\ref{fig:gap-t}.
The difference between both the cases depends on whether the radiative corrections with higher momentum quarks are dropped or not from the temperature effect.
In the low temperature and large chemical potential region, the phase boundaries of the first-order phase transitions for both the regularizations are almost identical, because the contribution is negligible from the momentum higher than the Fermi momentum.
The super restoration occurs only in the case 2.
It is found at about twice the temperature and chemical potential of the first-order phase transition and the peak of the chiral susceptibility.
Although the dimensions are different, the NJL model in the case 2 produces a similar behavior of the super restoration in the GN models.
The results of the GN models in two and three dimensions are independent of the regularization procedures because of their renormalizability.

\section{Summary and discussions} \label{sec:summary}
We have evaluated the massive four-fermion interaction models by using the effective potential in the leading order of the $1/N$ expansion under the assumption that the chiral condensate is spatially homogeneous.

First, we have considered the massive GN model on the $D$-dimensional spacetime ($2 \leq D < 4$).
At zero temperature and chemical potential, the chiral symmetry is broken above the critical coupling, $\lambda_r > \lambda_{\chi}$.
The dynamically generated fermion mass approaches the current mass, $m_r$, from above in the weak coupling limit, $\lambda_r \to 0$.
On the other hand, at a finite temperature and chemical potential, we have found the boundary where the dynamical fermion mass coincides with the current mass for a finite coupling, $\lambda_r$, on the $\mu$-$T$ plane as shown in Fig.~\ref{fig:boundaries} (with Figs.~\ref{fig:dynamical-mass:2dim} and \ref{fig:dynamical-mass:3dim}) based on Eq.~\eqref{eq:gap-equation:mass-chem-temp}.
We call this boundary super restoration boundary.
It is different from the well-known phase boundaries for the chiral symmetry.
The super restoration boundary is insensitive to changing the current mass, and the boundary at the massless limit gives a good approximation for a finite $m_r$ case.

Next, we have investigated the super restoration in the NJL model, a prototype model of QCD.
In four dimensions, the four-fermion models are non-renormalizable and the results depend on the regularization procedures.
We have employed two regularization procedures: the momentum cutoff is imposed to both the vacuum and thermal parts of the effective potential (case 1), and to only the vacuum part (case 2).
In the case 1, the dynamical fermion mass approaches but does not decrease below the current mass at high temperature and large chemical potential.
In the case 2, we have found the super restoration boundary where the dynamical fermion mass decreases across the value of the current mass.

In the models containing the explicit symmetry-breaking term (current quark masses), we have found the super restoration boundaries on the $\mu$-$T$ plane.
These boundaries represent the lines where the dynamical mass coincides the current mass.
On the boundaries, the spontaneously broken chiral symmetry is fully restored.
Outside the boundaries, the chiral condensate, $\Braket{\psibar \psi}$, develops a positive value (\eg $\Braket{\sigma} \simeq -(G/N_c) \Braket{\psibar \psi} < 0$), and the dynamical mass is smaller than the current mass.
The super restoration takes place at high temperature and large chemical potential region that is difficult to reach experimentally in QCD.

It has been pointed out that the inhomogeneous chiral condensate is favored at low temperature and large chemical potential \cite{Karasawa:2013zsa, Lenz:2020bxk}.
We will continue the work further and consider the inhomogeneous state.
We are also interested in applying our results to other systems and expect that the super restoration may be observed in some physical phenomena.
We hope to report on these problems in the future.

\section*{Acknowledgements}
The authors are indebted to members of our laboratory for encouragements and discussions.

\bibliography{references}
\end{document}